\documentclass[12pt,a4paper]{article}
\begin{document}

\title{Absorption spectrum of very low pressure, relatively cold atomic hydrogen.}
\author{Jacques Moret-Bailly\footnote{email: jmo@laposte.net}}
\maketitle
\begin{abstract}

 Impulsive, Stimulated Raman Scattering (ISRS) is a coherent light-matter interaction which may shift frequency of light while it propagates, as Rayleigh coherent scattering (refraction) dephases it. 2P hydrogen atoms pumped from 1S are able to stimulate ISRS of thermal, time-incoherent light if all involved time constants are longer than lengths of light pulses. Resulting redshift of light dilutes absorptions except if a previously absorbed line gets Lyman alpha frequency, stopping creation of 2P hydrogen. The redshift restarts slowly if energy at higher frequencies enables an absorption of other Lyman lines. The generated ``Lyman absorption forest'' generally starts and ends during a stop, so that total redshift is sum of fundamental redshifts which bring Lyman beta or gamma frequencies to Lyman alpha frequency. Karlsson's law is a good approximation of this result, remarking that the fundamental redshifts  are 3 or 4 times Karlsson's constant 0.061. 

\end{abstract}

Keywords:

290.5910 Scattering, stimulated Raman 

190.2640 Nonlinear optics : Stimulated scattering, modulation, etc.

\section{Introduction}
\label{Intro}
We study propagation of light emitted by an extremely hot thermal source, in very low pressure, relatively cold, pure atomic hydrogen (between 5 000 K and 50 000 K). We may suppose that, without optical excitation, the atoms are all free and in their ground state.

In his paper which founded the principle of lasers, Einstein \cite{Einstein1917} criticized implicitly his previous point of view on the quantization of the energy of light. He prepared quantum electrodynamics which splits the electromagnetic field into ``normal modes'' whose quanta of energy are  photons. W. E. Lamb, Jr., W. P. Schleich, M. O. Scully, C. H. Townes \cite{WLamb2} name the photon a pseudo-particle because the normal modes are relative to a choice which cannot be absolute because the definition of  normal modes, introduced comparing electrodynamics with acoustics, is that they do not exchange energy with other modes.

On the contrary, for most physicists, the photon is a small, elementary particle. Following Menzel \cite{Menzel}, in spite of his confusion between ``radiance'' and ``irradiance'', and of their use of refraction which is the {\it stimulated Rayleigh scattering}, the astrophysicists consider that coherent interactions are negligible.

 The quantum of energy detected at the focus of a telescope comes from a beam limited by the mirror and the diffraction spot of a far star. Using incoherent light, the mode of the photon, a propagating pulse of electromagnetic field, was around a nanosecond long, that is 30 cm thick. Far, it had a huge transverse size, it was a ``flat photon''.

The astrophysicists suppose that, in very low pressure of nebulae, their small photon is unable to interact simultaneously with several molecules. Thus, they assume ``on the spot interactions'' of light and molecules,  so that they study the propagation of light by inefficient, complex and arbitrary Monte-Carlo computations.

\medskip
 Coherence of a Raman scattering requires the same wavelength for exciting and scattered light. This may be obtained in optically anisotropic crystals, but in a gas it requires a broadening of the lineshapes by cutting light into pulses ``shorter than all involved time constants''(G. L. Lamb \cite{GLamb}).

\medskip
Show that using coherence of light-matter interactions is not only correct but also simple and powerful.

\section{Evaluation of variation of Impulsive Stimulated Raman Scattering (ISRS), according to the length of light pulses.}
Raman (quadrupolar) scattered light, made space-coherent by use of pulses, interferes with exciting light before appearance of flapping. An elementary Fourier computation shows that the main resulting frequency is intermediate between frequencies of both components, in proportion of their amplitudes.  The other components cancel by interferences. Thus, if Lamb's conditions are fulfilled,  Raman scattering shifts frequencies of light during its propagation, without blurring of images. Assuming, in a first approximation, that polarizability is independent on frequency, the relative frequency shift $\Delta \nu/\nu$ does not depend on $\nu$. It is the ``Impulsive Stimulated Raman Scattering'' (ISRS) \cite{Dhar}.

\medskip
Suppose that femtosecond pulses used in an ISRS experiment are replaced by $k$ times longer nanosecond pulses making time-incoherent light. How do the path needed for an observation of this ISRS is increased by the necessary increase by a factor $k$ of time constants ? To get only a coarse order of magnitude, we assume that pressure and resonance frequencies can remain optimal and other parameters constant.

- To multiply collisional time by $k$, pressure, thus ISRS shift, are decreased by factor $k$.

- Decreasing quadrupolar frequency by factor $k$,  Raman, thus ISRS frequency shift is divided by $k$.

- We have always a Stokes and an anti-Stokes scattering, whose results have opposite directions. Assuming thermal equilibrium, the difference of populations of quadrupolar levels, assumed weak, is proportional to quadrupolar frequency: ISRS frequency shift is divided by $k$.

\medskip
Thus, the order of magnitude of ISRS is reduced by a factor $k^3$, of order of $10^{15}$ : an observation of ISRS, easy  in a laboratory with 10 femtoseconds laser pulses, requires an astronomical path with incoherent light. 

\medskip
An ISRS tends to saturate a quadrupolar level, so that ISRS becomes weaker. Suppose that several ISRS are associated so that the populations of the quadrupolar levels remain constant: the gas becomes a catalyst, the interaction is ``parametric'', it exchanges energy between several light beams in accordance with thermodynamics.  The parametric effect includes always a contribution of the thermal radiation, which redshifts light. We name it Coherent Raman Effect acting on Incoherent Light (CREIL).

\section{Consequences of excitation of 1S atomic hydrogen by Lyman alpha absorption.}
The condition of coherence for an ISRS of incoherent light requires a quadrupolar resonance frequency lower than 1 GHz.

In its ground state, atomic hydrogen has the well known 1420 MHz quadrupolar resonance frequency, too high for an ISRS of incoherent light.

The 178 MHz in state 2S$_{1/2}$, 59 Mhz in  2P$_{1/2}$ and 24 MHz in 2P$_{3/2}$ are convenient. In more excited levels, the quadrupolar resonance frequencies work, but the ISRS is much weaker because the frequencies are too low.

\subsection{Lyman forest.}
Assume, for a simple theory, that a very hot source (star), emitting a {\it continuous spectrum} in the UV-X region is surrounded by an a light-year large region of decreasing pressure and temperature pure hydrogen.

All involved distances are much larger than the size of the source, so that the source may be considered as a point.

Close to the source, density and temperature decrease the free paths of atoms, they are assumed too large for an ISRS, so that the Lyman spectrum is strongly absorbed, written into light.

At a larger distance, the collisional time of atoms becomes larger than 1 nanosecond. The 1S atoms cannot be pumped to 2P because the Lyman lines were already absorbed. But assume that the temperature remains high enough to generate a few excited atoms, so that an ISRS redshifts slightly light. Frequencies slightly higher than Lyman $\alpha$ which were not absorbed, get the Lyman alpha frequency. Their absorption creates 2P atoms, the redshift increases, becomes fast; the redshift dilutes the absorption which becomes invisible, until the Lyman $\beta$ absorption line gets the Lyman $\alpha$ frequency $\nu_\alpha$: Assuming that the absorption of Lyman $\beta$ line was strong, the production of 2P hydrogen stops, the redshift stops.

As the redshift stops, all lines of the gas, mainly the Lyman absorption lines are strongly written into the spectrum. Thus, in the spectrum, we have two Lyman spectra, one shifted so that its $\beta$ line got the Lyman $\alpha$ frequency. 

If it remains in the light, energy at frequencies high enough to excite the 1S atoms to levels higher than 2P, a weak redshift is produced by atoms excited to high levels, or by 2S or 2P atoms resulting from a cascade from high levels. The weak redshifting brings unabsorbed frequencies at the Ly$\alpha$ frequency, a fast redshift works until {\it any} absorbed line gets Lyman alpha frequency. Each absorbed line generates new absorbed lines, the complexity of spectrum increases from low frequencies to higher.

Thus we obtain sets of Lyman absorptions shifted several times by the fundamental shifts which bring frequencies $\nu_\beta$ or $\nu_\gamma$ to  $ \nu_\alpha$ frequency (assuming that Ly$_\delta$ absorption is negligible). This spectrum is a ``Lyman absorption forest'' \cite{Rauch}.

Having assumed that the source is relatively small, space is divided into regions (shells) which redshift light, separated by regions which do not. 

The cycles of shifts and absorptions have a large probability to stop during an absorption phase because an exit from this phase requires pumping at high frequencies while Planck's formula shows, at high frequencies, a fast decrease of radiance of a thermal source by increase of frequency: The irradiance was high enough to shift the spectrum to the stop, a much larger irradiance at higher frequency would be necessary to restart.

\medskip
Simple theory assumes that gas has only three spectral lines, Lyman $\alpha, \beta$ and $\gamma$, and that the radiance of the source varies slowly. The spectrum of a real thermal source shows many, generally weak lines, in particular absorption lines between Ly$_\alpha$ and Ly$_\beta$. Some, among these lines, may slow redshift down before the first coincidence of absorbed Ly$_\alpha$ line with Ly$_\beta$ line, filling a spectral region left virgin by the simple theory, with more or less strongly absorbed lines.

\subsection{Karlsson's formula.}

Set $Z_{(\nu_0,\nu_1)}=(\nu_0-\nu_1)/\nu_0)$ a redshifts which brings an initial frequency $\nu_0$ to frequency $\nu_1$.

As the cycle of redshifts and absorptions starts and stops generally during an absorption, using the simple theory, the largest observed redshift is generally close to $bZ_{(\alpha, \beta)}+cZ_{(\alpha, \gamma)}$, where $b$ and $c$ are non-negative integers and  $Z_{(\alpha, \beta)}$ (resp.$Z_{(\alpha, \gamma)}$) is the redshift which transforms Ly$_\beta$ (resp. Ly$_\gamma$ ) frequency into Ly$_\alpha$ frequency. By Rydberg's formula:

$Z_{(\beta,\alpha)} = (\nu_\beta-\nu_\alpha)/\nu_\alpha = [(1-1/3^2 -(1-1/2^2)]/(1-1/2^2) ] \approx 5/27 \approx 0.1852 \approx 3*0.0617; $ (1)
 	
$Z_{(\gamma,\alpha)} = (\nu_\gamma-\nu_\alpha)/\nu_\alpha = [(1-1/4^2 -(1-1/2^2)]/(1-1/2^2) ] = 1/4 = 0,25 = 4*0.0625$; (2)

The largest redshift (redshift of the star) is generally close to the product of Karlsson's constant $K = 0.061 $ by an integer $q$ sum of integers 3 and 4.
There is an overlap of lines, so we build a tree whose branches may merge. Certain values of q are remarkable, eg $ q = 10$ corresponds to different combinations of redshifts ($10= 3 +3 +4 = 3 +4 +3 = 4 +3 +3 $), it corresponds to a large probability of an observed largest redshift.

Absorption and redshifts occur in well defined regions because the source of light is small. If the source of light is large , these regions cannot be defined, Karlsson's formula does not work.

\section{Conclusion.}
Considering that the photon is a pseudo-particle resulting from the quantization of ``normal modes'' of the electromagnetic field allows the use of effects used in laser spectroscopy such as the Impulsive Stimulated Raman Scattering, with unusual space-time parameters.
We study by theory propagation of light emitted by an extremely hot thermal source, in very low pressure, relatively cold atomic hydrogen.
We take into account the formation of 2P hydrogen which catalyzes a parametric exchange of energy between light beams of time-incoherent light, in accordance with thermodynamics. Thus frequencies are shifted during propagation of light, without any blurring of images.

Standard astrophysical theories explain Lyman forests assuming existence of filaments of hydrogen, without any justification of their structure by independent observations.
The redshifts of quasars and nearby, compact galaxies are generally close  to prevalent values obtained by statistics \cite{Karlsson,Burbidge}. This result and empirical Karlsson's formula which gives the prevalent values do not have any standard explanation. Our elementary spectroscopy explains the existence of prevalent frequency shifts, provides both parameters of Karlsson's formula and explains why it does not apply to big galaxies.

The computation which provides Karlsson's formula provides also a simplified spectrum of ``Lyman forest'' of quasars because it uses only the three most intense lines of H atom. Other lines of H and other atoms may be easily introduced, retaining the simple principle of theory: There is no need to introduce hydrogen filaments. The low density of lines observed close to the Ly $\alpha$ emission line (proximity effect) is a consequence of the multiplication of the lines from the start of our process. These general similarities with the quasar spectra \cite{Rauch}. allow us to assert that we have the key of a precise nomenclature of the lines of the quasar spectra.

The use of ISRS and the combination of ISRS making CREIL seems an alternative of big bang which, as refraction, depends on a dispersion, so that it explains that the emission multiplets of the quasars are distorted. Astrophysicists should carefully study an introduction of ISRS and, more generally, optical coherence, among their tools.

\end{document}